\documentclass[12pt,preprint]{aastex} 
\usepackage{amssymb}
\usepackage{amsmath}
\usepackage{graphicx}

\newcommand{\dist}{h^{-1}\,{\rm Mpc}}
\newcommand{\mass}{h^{-1}\,{\rm M_{\odot}}}

\shorttitle{Supercluster Shape and Evolution}
\shortauthors{Wray et al.}
\begin{document}

\title{The Shape, Multiplicity, and Evolution of Superclusters in $\Lambda$CDM
 Cosmology}
\author{James J. Wray\altaffilmark{1}}
\email{jwray@astro.princeton.edu}
\author{Neta Bahcall\altaffilmark{1}}
\email{neta@astro.princeton.edu}
\author{Paul Bode\altaffilmark{1}}
\email{bode@astro.princeton.edu}
\author{Carl Boettiger\altaffilmark{1}}
\email{cboettig@princeton.edu}
\author{Philip F. Hopkins\altaffilmark{2}}
\email{phopkins@cfa.harvard.edu}

\altaffiltext{1}{Princeton University Observatory, Princeton, NJ 08544, USA}
\altaffiltext{2}{Harvard-Smithsonian Center for Astrophysics,
 60 Garden Street, Cambridge, MA 02138, USA}

\begin{abstract}
 We determine the shape, multiplicity, size, and radial structure of
 superclusters in the $\Lambda$CDM concordance cosmology from z = 0 to z = 2.
 Superclusters are defined as clusters of clusters in our large-scale
 cosmological simulation.  We find that superclusters are triaxial in
 shape; many have flattened since early times to become nearly
 two-dimensional structures at present, with a small fraction of filamentary
 systems.  The size and multiplicity functions are presented at different
 redshifts.  Supercluster sizes extend to scales of $\sim100 - 200\dist$.
 The supercluster multiplicity (richness) increases linearly with supercluster
 size.  The density profile in superclusters is approximately isothermal
 ($\sim R^{-2}$) and steepens on larger scales.  These results can be used as
 a new test of the current cosmology when compared with upcoming observations
 of large-scale surveys.
\end{abstract}


\clearpage

\section{INTRODUCTION}
 Superclusters are the largest structures in the Universe.
 Their sizes span the range from a few times the radius of a typical
 galaxy cluster ($\sim$ 3-4 $\dist$) to the
 lengths of the ``Great Walls'' of galaxies observed by the CfA Redshift
 Survey \citep{geller89} and the Sloan Digital Sky Survey \citep{gott05}.
 These latter sizes of $\sim10^2 \dist$ represent a significant fraction
 ($\sim 10$\%) of the horizon scale.

 Because they are so large relative to a typical survey volume, superclusters
 have proven difficult to study, especially in a statistical manner.
 Observationally, attempts to characterize the large-scale structure using
 distributions of galaxies date back to \citet{hubble36}, but these methods
 did not become powerful until the advent of galaxy cluster and galaxy
 redshift surveys.
 \citet{abell58} used the Palomar Sky Survey to construct a catalog of
 rich clusters, with which he was able to find evidence for what he called
 ``second-order clusters,'' i.e. clusters of clusters, which are indeed
 superclusters.  Subsequent work included \citet{gregory78}'s study
 of the supercluster containing the rich Coma cluster, and
 \citet{gregory81}'s work on the Perseus supercluster.
 Other observational approaches to supercluster identification have
 been based, like that of \citet{abell58}, on cluster catalogs
 \citep[e.g.][]{oort83, bahcall84, bahcall88, einasto94, kolokotronis02} or
 on analysis of smoothed density fields 
 \citep{einasto03a,einasto03b,erdogdu04,einasto06}.  These have
 been important in characterizing the low-redshift Universe, but they
 have not yet allowed the study of supercluster evolution from early
 cosmic times (i.e., from redshifts of order unity).  How do superclusters
 form, and how do their shapes, multiplicities, sizes, and structures
 evolve with time?  The answers to these fundamental questions are not yet
 known.

 Recently, there has been exciting progress on both the
 observational and computational fronts.  The Sloan Digital Sky Survey
 \citep[SDSS,][]{york00, stoughton02} is collecting deep, broadband
 imaging and spectroscopy of a large, contiguous portion of the sky.  SDSS
 has created the largest galaxy redshift survey catalog to date, and much
 work is also underway to improve photometric redshift techniques
 \citep[e.g.][]{csabai00, budavari00}, which would permit measurements of
 3-D galaxy positions for the much larger SDSS imaging catalog.
 On the theoretical side, a concordance cosmological model has gained
 widespread acceptance \citep{bahcall99, bennett03, spergel03, spergel06}.
 Applying this model to cosmological dynamics codes such as that described
 below, we can now reliably predict the properties and
 evolution of superclusters.  The time is therefore ideal for a
 computational analysis of large-scale structure using new, powerful
 simulations, the results of which can be 
 used to determine the properties and evolution
 of superclusters.
 In addition, these results can be compared with observations, thus
 providing a new test of the
 current cosmological model---complementary to that provided by CMB,
 large-scale power spectra, and other
 studies---as well as shed light on structure formation and evolution.

 We begin by describing, in \S2, the process by which clusters and in turn
 superclusters were identified in the simulation output.  We provide
 quantitative descriptions of the multiplicity and size distributions of
 the identified superclusters
 in \S\ref{sec:mult} and \S\ref{sec:size}. 
 We examine the radial structure of our
 superclusters in \S\ref{sec:radial} and  the dimensionality of
 superclusters in \S\ref{sec:shapes}.  
 The evolution of the above properties with redshift
 is presented in \S\ref{sec:evo}.  
 Results for superclusters selected from only higher-mass
 clusters are presented in \S\ref{sec:him}.  
 Our conclusions are summarized in \S\ref{sec:conc}.

\section{SELECTION OF SUPERCLUSTERS}
\subsection{The Simulation}
 We use the efficient, parallel dark matter simulation code
 of \citet{bode03}.  A large-scale, high-resolution
 simulation was evolved with a particle-mesh method to 
 compute long-range gravitational forces, and a tree code to
 treat high-density regions.  We use the cosmological parameters
 of the concordance model: $\Omega_{m}=0.27, \Omega_{\Lambda}=0.73,
 {\rm H}_{0}=70\,{\rm km\,s^{-1}\,Mpc^{-1}}, n_s=0.96,$ and
 $\sigma_{8}=0.84$ 
 (cf. \citet{spergel03}; see also \citet{spergel06}).
 A total of $1260^3$ particles, each assigned the mass
 $1.264\times10^{11} \mass$, is evolved in a periodic box $1500 \dist$ on
 a side, and positions are saved at the appropriate times in order to
 construct a light cone with vertex at one corner of the box.
 Clusters are then identified in two different,
 partially overlapping volumes: a ``low-redshift'' sphere of radius
 $1500 \dist$, corresponding to a maximum redshift of $z \approx 0.57$,
 and a ``high-redshift cone,'' which includes the low-$z$
 sphere's positive octant ($x > 0, y > 0, z > 0$) and extends this octant
 out to $z = 3$, a distance of $\sim4600 \dist$.  Using the cluster
 selection scheme described below, a total of 1,442,616
 clusters  with $M_{vir} \ge 1.75\times10^{13} \mass$
 are found in the low-$z$ sphere, and 720,550 in the high-$z$
 cone.  Further details on this light cone simulation are given
 by \citet{hopkins05}, hereafter HBB.

\subsection{Cluster Selection}
 Clusters are identified in the simulation using the friends-of-friends
 (FOF) algorithm \citep[see HBB;][]{davis85}.  The FOF results depend on
 the chosen linking length $L$.  A common parameterization uses the
 linkage parameter $b \equiv L/r_{av}$, where $r_{av}$ is the mean
 interparticle separation.  This is useful in applications where the mean
 density of objects, and in turn their mean separation, changes over time.
 Holding $b$ constant imposes the same minimum overdensity for clusters
 at all redshifts.  We use the value $b = 0.2$ (see also HBB), corresponding
 to an overdensity of $\sim180$ times the mean density.  The center of each
 FOF cluster is taken to be the position of the most bound particle.  This center
 is used to compute the virial radius $R_{vir}$, defined as the radius enclosing
 the virial overdensity expected from spherical top-hat collapse (which evolves
 with redshift in the $\Lambda$CDM cosmology).  In this analysis,
 we consider only clusters with virial mass threshold (i.e., the total mass
 within the virial radius) $M_{vir} \ge 1.75\times10^{13} \mass$.  This mass
 threshold is typical for a poor cluster of galaxies.  The mean mass of all
 clusters in the low-$z$ sphere is $\sim4.40\times10^{13} \mass$.  In 
 \S\ref{sec:him}
 we discuss superclusters selected from clusters with a higher mass
 threshold.

 The ellipticities of the clusters identified at low and high $z$ are
 studied in detail by HBB.  Relevant to our present work on
 superclusters, they find alignments between nearby cluster pairs out to
 separations of ${\sim}100 \dist$; the strength of alignment increases
 with redshift and with decreasing cluster separation.  HBB explain these
 alignments as being due to clusters forming and gathering material
 along ``filamentary superstructures''; they provide evidence for
 overdensities along the separation vectors between pairs of aligned
 clusters.

\subsection{Supercluster Selection}
 As demonstrated by \citet{abell58}, \citet{bahcall88} and references
 therein, superclusters can be identified by their overdensity of
 clusters.  We use the FOF method to identify superclusters from the
 cluster sample defined above (\S2.1-2.2).  The mean density of
 clusters in the low-$z$ sphere (we shall not discuss the high-$z$ cone
 until \S\ref{sec:evo}) 
 is $n_{av} = 1.02\times10^{-4} (\dist)^{-3}$, yielding a mean
 intercluster separation of $r_{av} \equiv n_{av}^{-1/3} = 21.4 \dist$.
 We therefore select linking lengths of $L = 3, 5, 7,$ and $10 \dist$,
 corresponding to $b \approx 0.14, 0.23, 0.33,$ and $0.47$, respectively.  
 All scales in this paper are in comoving coordinates.
 The different linking lengths sample different portions of the
 supercluster geometry.  Small linking lengths identify only the dense
 cores of superclusters, whereas larger linking lengths allow the FOF
 algorithm to percolate to lower-density filaments of clusters.  Large
 $L$-values therefore reach to the outer parts of superclusters, as well
 as identify looser superclusters, and often connect structures that
 appear discrete with smaller $L$.

 We use Monte Carlo simulations to compare the significance of the
 selected superclusters with that of random associations.  Clusters are
 placed at randomly chosen positions within a sphere such that the total
 number density is the same as that for the simulated clusters, and the FOF
 method is applied, as before, with $L = 3, 5, 7,$ and $10 \dist$.  We find
 significantly fewer clusters per supercluster, and significantly fewer
 superclusters down to multiplicity as small as 3, for all four linking
 lengths tested.  This result implies that most of the superclusters are real,
 even for $L = 10 \dist$.  The only exception is the case of binary clusters
 (superclusters with only 2 members) selected with $L = 10 \dist$, of which
 we find only $\sim25$\% more using the simulated clusters than we do with
 the Monte Carlo sample.  The random rate of superclusters is typically
 $<10$\% for $N>3$ superclusters, decreasing to $<1$\% for $N>6$
 superclusters.  The Monte Carlo results are compared with the $\Lambda$CDM
 multiplicity functions below.

\begin{figure*}
\resizebox{0.75\hsize}{!}{\includegraphics[angle=270]{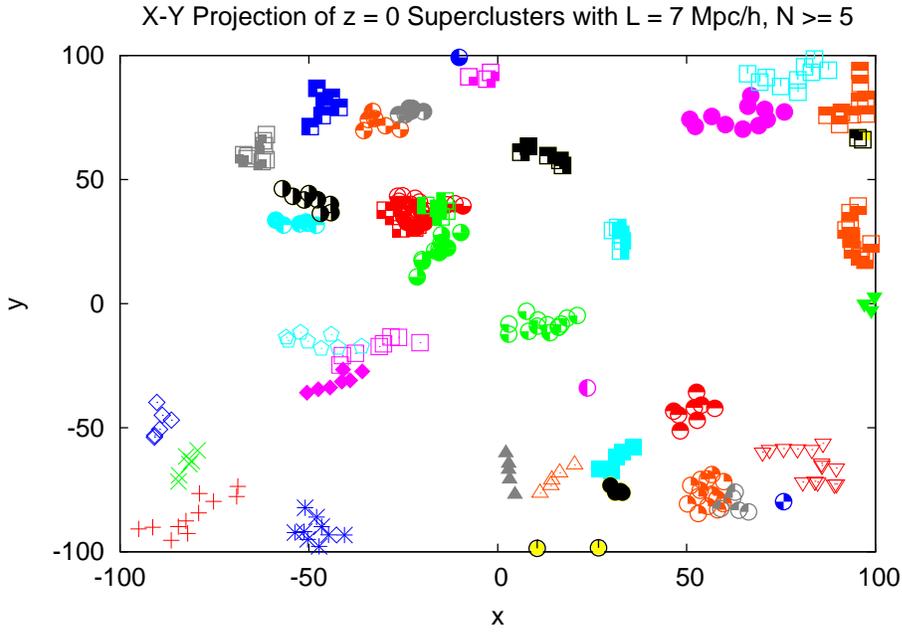}}
\caption{\label{fig1}
 Clusters identified as supercluster members at redshift zero, for
 $L = 7 \dist$ and minimum multiplicity 5.  Clusters represented by similar
 points are members of the same supercluster.  See the text (\S2.3) for
 further description.
}
\end{figure*}

\begin{figure*}
\resizebox{0.75\hsize}{!}{\includegraphics[angle=270]{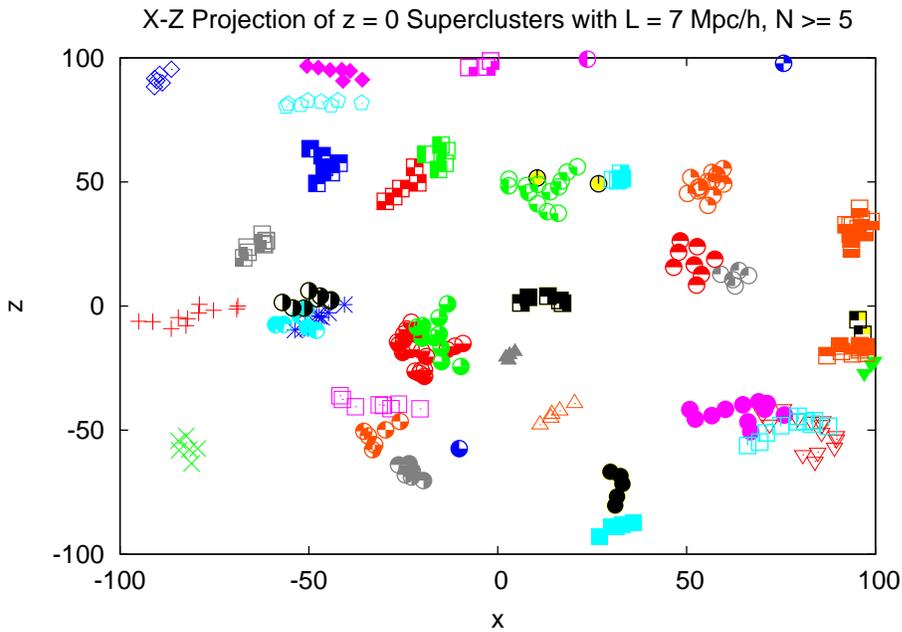}}
\caption{\label{fig2}
 Same parameters as Fig$.$\,\ref{fig1}, but a different projection.
}
\end{figure*}

 To help visualize the typical geometries of superclusters identified
 by the FOF method, we present supercluster maps of a typical region (a cube
 of side length $200 \dist$, concentric with the low-$z$ sphere) for
 $L = 7 \dist$ (Figs$.$\,\ref{fig1}-\ref{fig2}).  
 Each point is a cluster, and a given symbol
 represents membership in the same supercluster.  The maps present
 a projection of the 3-D distribution onto the $xy$-plane in the first
 plot, and onto the $xz$-plane in the second plot.  Viewing the
 plots in turn corresponds to a $90^{\circ}$ rotation of the viewing angle
 about the $x$-axis.  For clarity, we show only superclusters with at least
 5 member clusters.  Members of a supercluster that extend outside the limits
 of our cubical region are not shown, so some superclusters appear to have
 fewer members than the imposed minimum.  Comparing these maps to analogous
 figures generated with $L = 5$ and $10 \dist$ reveals, as expected, that
 increasing the linking length often causes neighboring
 superclusters to be connected into one.  This is especially true when $L$
 increases from 7 to $10 \dist$, resulting in many large, complicated
 structures with central cores joining multiple lower-density filaments.

\section{SUPERCLUSTER PROPERTIES}

\subsection{Multiplicity Function}     \label{sec:mult}
\begin{figure*}
\resizebox{\hsize}{!}{\includegraphics[angle=0]{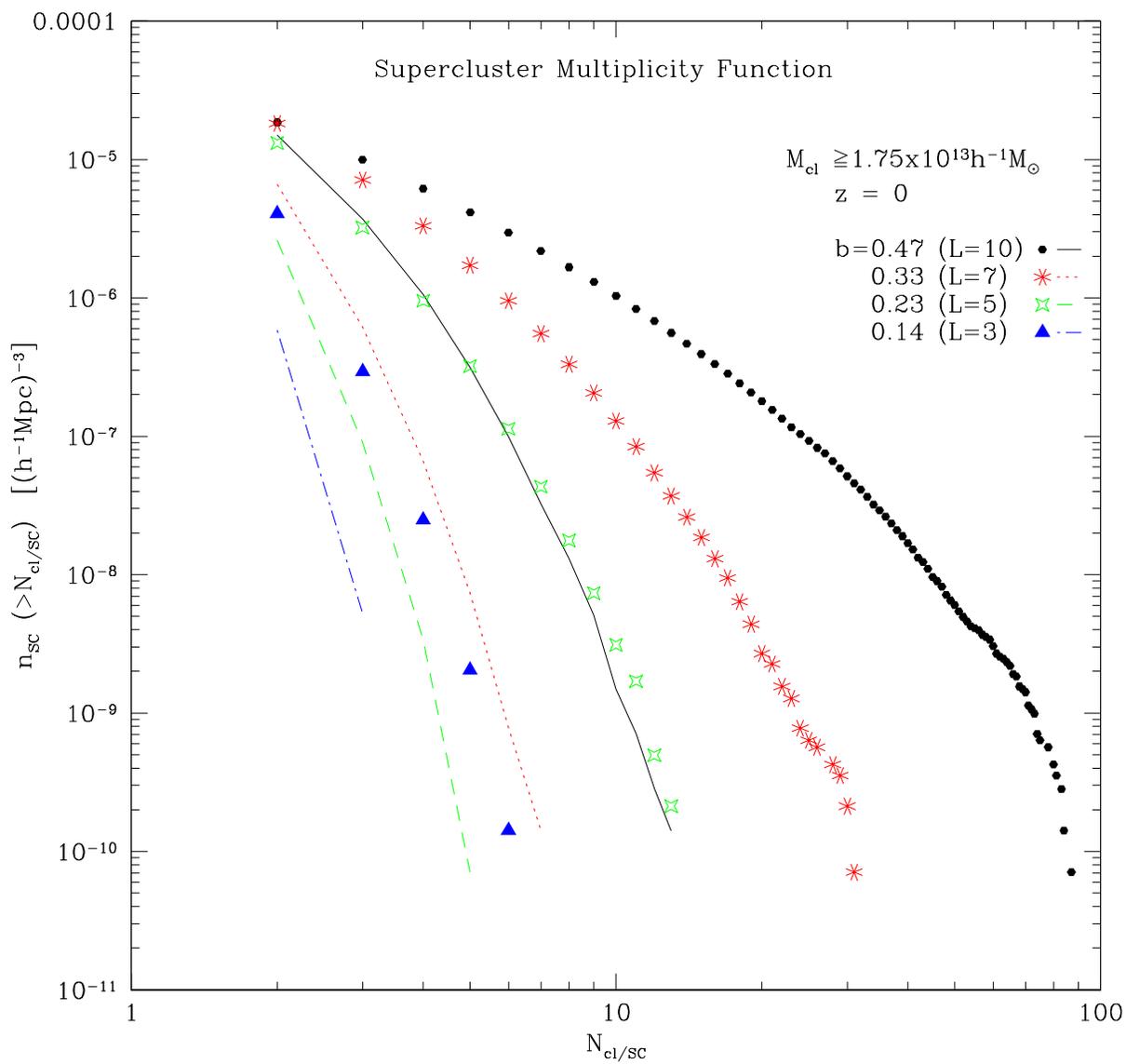}}
\caption{\label{fig3}
 Integrated supercluster multiplicity functions for linking lengths $L = 3, 5, 7,$
 and $10 \dist$, all at $z \approx 0$.  Curves show superclusters identified in our
 random Monte Carlo simulations to test the physical reality of the superclusters.
}
\end{figure*}

 The {\em multiplicity} of a supercluster is the number of clusters
 it contains, $N$.  In Fig$.$\,\ref{fig3} 
 we present, for various linking lengths, 
 the integrated supercluster multiplicity function --- the number density
 (per cubic comoving $\dist$) of superclusters with $N$ or more members, as a
 function of $N$.   Also shown for comparison are the corresponding
 results from random catalogs, as described in the previous section.

 Longer linking lengths yield more superclusters of all
 multiplicities, as expected.  Shorter linking lengths yield
 curves that rise more steeply toward low multiplicities, implying that
 they preferentially find small superclusters with low multiplicity.
 Conversely, longer linking lengths find a higher percentage of larger,
 richer superclusters.  The shorter linking lengths naturally
 trace high-density regions (e.g., dense cores of superclusters), while
 larger linking lengths characterize lower-density regions, such as the
 outer parts of superclusters or diffuse superclusters.
 Small $L$ can also divide structures that are identified with larger
 linking lengths.  The fact that we see superclusters with $\sim40$ or more
 clusters may seem surprising, but the total number density of such
 structures is only $10^{-8}$ per $(\dist)^3$, and it is based on a very low
 mass threshold of cluster members, $M_{vir} \ge 1.75\times10^{13} \mass$.
 Such clusters are numerous, and any comparison with observations should of
 course account for this threshold.  For superclusters selected using
 higher-mass clusters, see \S\ref{sec:mevo}.
 We repeated the above analysis after selecting superclusters 
 in redshift-space coordinates; the resulting multiplicity
 functions are nearly identical to the real-space functions at 
 low multiplicities, and are shifted to slightly higher $N$ 
 (by up to 20\%) at the highest multiplicities.

\subsection{Size Function}     \label{sec:size}
\begin{figure*}
\resizebox{\hsize}{!}{\includegraphics[angle=0]{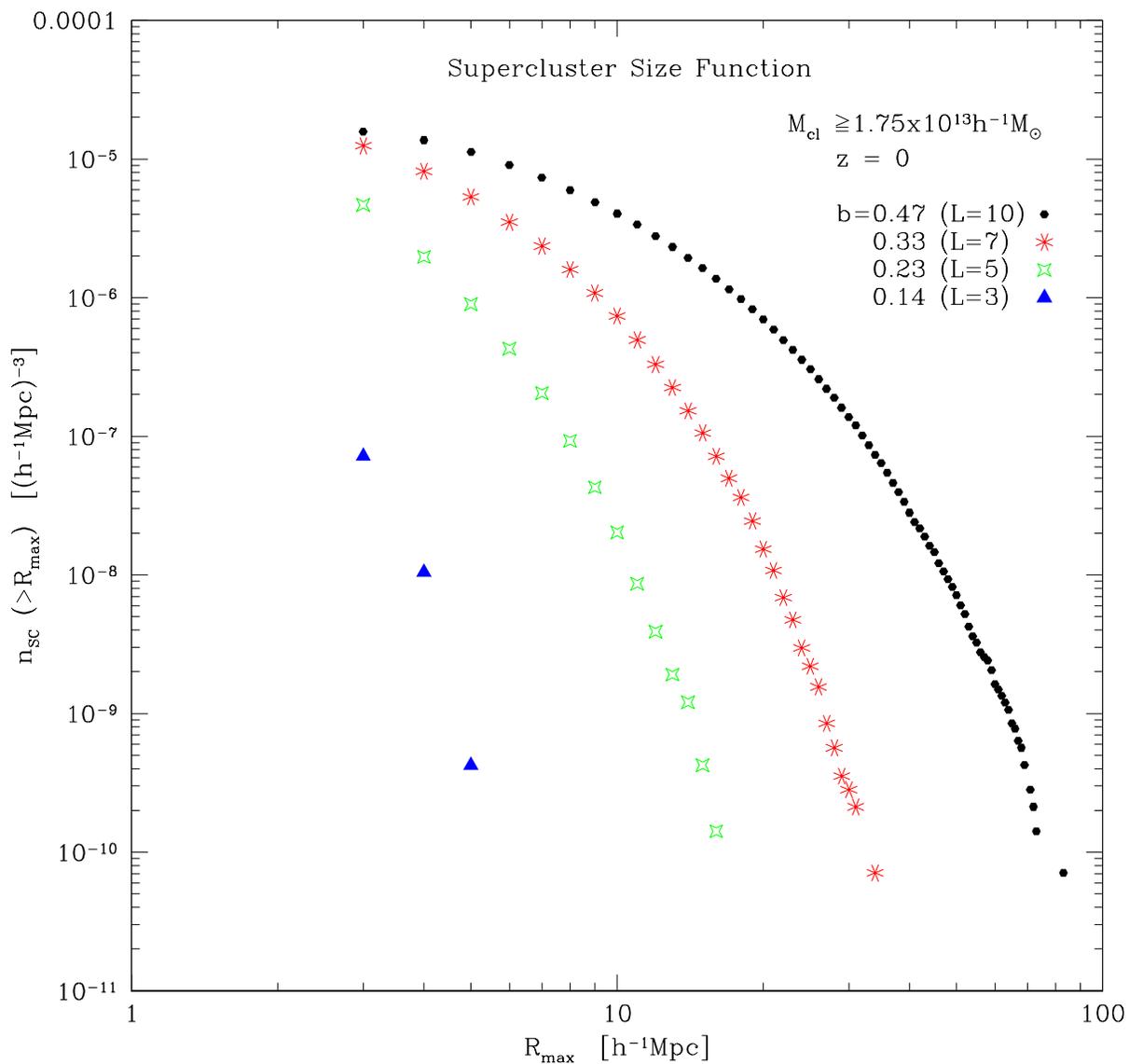}}
\caption{\label{fig4}
 Integrated size functions for linking lengths $L = 3, 5, 7,$ and
 $10 \dist$ at $z \approx 0$.  The size measurement along the horizontal
 axis is, for a given supercluster, the distance of the member cluster
 most distant from the supercluster center of mass.  All are comoving scales.
}
\end{figure*}

 The supercluster size function --- i.e., the number of superclusters above
 a given size --- is presented in Fig$.$\,\ref{fig4}
 for different linking lengths.  The
 size of a supercluster is defined as the maximum distance of any member
 cluster from the supercluster center of mass.  As with the multiplicity
 function, the size function rises more steeply for shorter linking lengths.
 This implies that the longer linking lengths find a greater ratio of large
 superclusters to small superclusters, analogous to the result for the
 multiplicity function. 
 Similar results are found using redshift-space coordinates;
 for the largest linking length used, the sizes are slightly
 ($\lesssim 5\%$) larger.

 We find that the largest supercluster radii are $\sim 80 \dist$,
 corresponding to diameters of $\sim160 \dist$.  This is
 consistent with the largest structures seen in observations
 \citep[e.g.][]{bahcall84,geller89,gott05}, and it is also the
 largest distance at which HBB detect cluster alignment in the
 simulation.

\subsection{Radial Structure}     \label{sec:radial}
\begin{figure*}
\resizebox{\hsize}{!}{\includegraphics[angle=0]{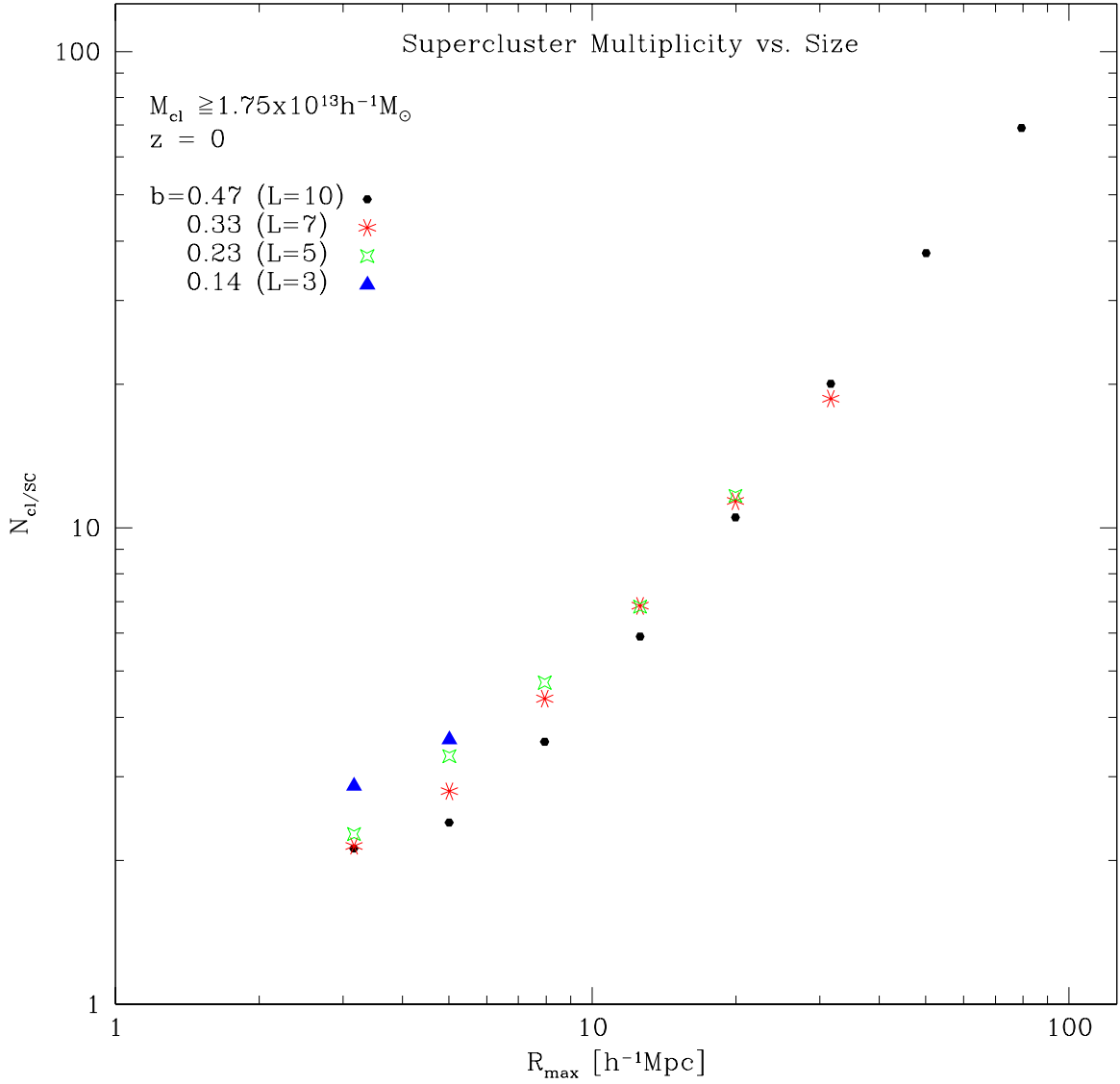}}
\caption{\label{fig5}
 Multiplicity vs. size of superclusters, $z \approx 0$.  $R_{max}$ is
 defined as in Fig$.$\,\ref{fig4}.  Linking lengths $L$ are quoted in $\dist$.
}
\end{figure*}

 The average multiplicity of superclusters is plotted as a function of
 supercluster size for different linking lengths in Fig$.$\,\ref{fig5}.
 For $N = 2$
 or $3$ --- objects more aptly called binary or triple clusters than
 superclusters --- the maximum radius is set mainly by the
 separation(s) of the 2 or 3 clusters; thus the average
 multiplicity changes slowly over this range of separations.  Beyond
 this regime, the multiplicity and size of superclusters appear to be
 related nearly linearly.  This result suggests that
 the spherically averaged density profile in superclusters is roughly {\em 
 isothermal}; i.e., $n(r) \propto r^{-2}$, where the cluster number density
 $n(r)$ satisfies:
 $$
 N = \int_{0}^{R_{max}} n(r)4{\pi}r^2{\rm d}r
 \eqno(1)
 $$
 This profile yields multiplicity $N \propto R_{max}$ for
 $n(r) \propto r^{-2}$.

 We can investigate the spherically averaged density profile more directly
 by examining the distribution of clusters within
 individual superclusters and determining the average number
 density profile for the entire sample.  To do so, we superpose
 the centers-of-mass (calculated using only the mass in clusters) of all
 superclusters, and scale all superclusters to the same size by dividing
 all cluster coordinates (relative to the center-of-mass) for a given
 supercluster by its $R_{max}$.  We add up the total number of clusters
  inside radial bins, divided by the total
 number of clusters in superclusters for the given linking length.
 We have included
 only superclusters with 5 or more members;
 because the multiplicity function drops off rapidly,
 most of the superclusters included have only 5 or 6
 members.  This fact affects the combined profile in two ways.  First,
 only in few cases will one of the five members happen to lie very
 near the supercluster's center-of-mass; thus it is only at
 $R/R_{max} \gtrsim 0.2$ that we begin to find a significant fraction of
 the clusters.  Second, one out of roughly five clusters in each
 supercluster lies at $R/R_{max} = 1$.  Thus, for the smaller
 linking lengths, the integrated number fraction approaches a limit of
 $\sim 0.8$; for the larger linking lengths, which find richer
 superclusters, the limiting fraction is slightly higher.  The fraction
 of clusters within $R/R_{max}$ increases linearly with radius for the
 range $0.3 \lesssim R/R_{max} \lesssim 0.7$.  This linear
 relationship indeed implies $n(R) \sim R^{-2}$ for the supercluster
 density profile in this range.  The profile steepens on larger scales.

 The average mass of clusters in superclusters,
 $M = 4.8\times10^{13} \mass$ for the $L = 7 \dist$ superclusters,
 is somewhat larger than the average mass of clusters in the total sample,
 $M = 4.4\times10^{13} \mass$, indicating that superclusters contain slightly
 more massive clusters, on average, than the mean.  The $L = 7 \dist$
 superclusters with $N \ge 5$ members have an even higher average cluster
 mass of $M = 5.5\times10^{13} \mass$, implying that richer superclusters
 contain more massive clusters than poorer superclusters.

\subsection{Supercluster Shapes} \label{sec:shapes}
 We measure the shape, and in particular the dimensionality, of a
 given supercluster by fitting a 3-D ellipsoid to the distribution of its
 member clusters.  We use the code developed by HBB to
 fit ellipsoids to the dark matter particles comprising their clusters.
 The code constructs, for a given supercluster, the $3\times3$ matrix of
 second moments of cluster positions relative to the supercluster
 center-of-mass, i.e.
 $$
 I_{ij}=\sum x_{i}x_{j}m,
 \eqno(2)
 $$
 where $x_{i}$ and $x_{j}$ are two of the position coordinates for a given
 cluster, and $m$ is the mass of the cluster; the sum is over all member
 clusters.  The eigenvalues of this matrix are simply the three axis lengths
 of the best-fit ellipsoid for the supercluster, times a known constant.
 Following the notation used by HBB, we shall denote these lengths
 $a_1, a_2, a_3$ for the primary,
 secondary, tertiary axes, respectively; i.e., $a_1 \ge a_2 \ge a_3$.
 We restrict our analysis to superclusters with $N \ge 5$; for
 poorer superclusters, the ``best-fit ellipsoid'' would not be very
 meaningful.

 We have already examined size distributions for superclusters 
 (\S\ref{sec:size}), so
 we now focus on the {\em axis ratios}, which allow us to probe the shape
 and dimensionality of superclusters.  Fig$.$\,\ref{fig6}
 displays the distribution of
 {\em primary axis ratios} $a_2/a_1$ for superclusters selected with
 $L = 5,7,10 \dist$.  Fig$.$\,\ref{fig7}
 shows the distribution of {\em secondary axis ratios} $a_3/a_2$,
 and Fig$.$\,\ref{fig8} shows $a_3/a_1$.  
 The mean and peak axis ratios from each curve in these
 figures are listed in Table 1.

 The secondary axis ratios are generally smaller than the primary
 axis ratios.  We find that a primary axis ratio of $a_2/a_1 \approx 0.6$ is
 typical for superclusters selected using a broad range of linking
 lengths; this corresponds to a primary ellipticity of
 ${\epsilon}_1 \equiv 1-a_2/a_1 \approx 0.4$.  The secondary axis ratio
 is more strongly dependent on the linking length used to
 find the superclusters.  For $L = 5 \dist$, we find a peak $a_3/a_2$ of
 $\sim 0.4$, corresponding to a secondary ellipticity of
 ${\epsilon}_2 \equiv 1-a_3/a_2 \approx 0.6$; for $L = 10 \dist$, we 
 find a peak of $a_3/a_2 \approx 0.2$, or ${\epsilon}_2 \approx 0.8$.
 The peak $a_3/a_1$ ratios range from $\approx 0.13$ for
 $L = 10 \dist$ to $\approx 0.23$ for $L = 5 \dist$.

 These measurements allow us to construct a picture of the typical
 shape and dimensionality of superclusters.  The interpretation of
 different ellipticities can be described as follows:
 \begin{enumerate}
 \item A combination of 
    high primary and low secondary axis ratios is typical of
    two-dimensional pancake-like structures (two dimensions are large,
    and one is small).
 \item A combination of low primary and high secondary axis ratios is
    typical of one-dimensional filamentary structures (one dimension
    is large, and two are small).
 \item A combination of primary and secondary axis ratios that are both
    significantly less than unity is
    typical of structures most appropriately described as triaxial
    (each dimension is a different size).
 \item If both axis ratios are close to unity, then the distribution is nearly
    spherical.
 \end{enumerate}
 The results we find suggest that superclusters are typically triaxial
 structures nearing 2-D, pancake-like structures at $z\approx0$.  They
 are neither spherical nor purely filamentary in nature.  As the linking
 length is increased, the resulting superclusters become more two-dimensional;
 the cores of superclusters are thus more triaxial, spreading out to more
 pancake-like structures on larger scales.

 We plot the bivariate distribution of $a_2/a_1$ and $a_3/a_2$ 
 for $L = 7 \dist$ in Fig$.$\,\ref{fig9}
 (the distributions for $L = 10$ and $5 \dist$ are similar).
 Applying the interpretation rules
 listed above, 1-D structures reside
 in the upper left corner of Fig$.$\,\ref{fig9}, 2-D in the lower right
 corner, and 3-D (spherical) in the upper right.  Points in the middle
 or toward the lower left corner represent triaxial structures.

 Fig$.$\,\ref{fig9}
 shows that a large majority of the superclusters
 are more nearly two-dimensional than one-dimensional.
 No nearly spherical
 superclusters are found with this linking length, but there is a smaller
 population of filamentary structures.  
 Similar distributions are found for $L = 10$ and $5 \dist$, though
 showing more flattened structures (2D) as $L$ increases,
 and more triaxial structures as $L$ decreases.
 This is the same trend depicted in Fig$.$\,\ref{fig7}
 as a shift toward lower
 secondary axis ratio with increasing linking length.  
 The shorter linking lengths also
 yield significantly higher fraction of filamentary structures --- i.e.,
 superclusters with low primary axis ratio and a secondary axis ratio
 approaching unity.

 This trend suggests that most superclusters at $z \approx0 $
 are triaxial structures with one dimension considerably smaller than
 the other two.
 This geometry is apparent in Fig$.$\,\ref{fig9}, and becomes more
 significant when a larger FOF linking
 length is used.  Since the superclusters are not completely flat,
 smaller linking lengths may probe only
 the dense core regions, which are smaller in size than the
 supercluster thickness.  Thus, while the thickness sets an upper bound
 to the tertiary axis length identified with any linking length, the
 higher $L$-values find longer primary and secondary axes by probing
 farther out into the pancake.  This causes larger linking lengths to find
 a smaller mean secondary axis ratio, and reveal the underlying 2-D nature
 of the large-scale superclusters.  Similarly, for a given $L$, the largest,
 highest-multiplicity superclusters ($N>10$) are on average flatter
 ($a_3/a_2\approx0.24$ for $L=10\dist$) than those  with $N=$ 5-10
 ($a_3/a_2\approx0.35$).
 This flatness is still apparent in projection:
 for superclusters
 selected (from $M_{vir} \ge 1.75\times10^{13} \mass$ clusters)
 in redshift-space, the projected axis ratios have a broad
 distribution with a mean $a_2/a_1=0.4\pm0.005$
 and a peak $a_2/a_1\approx0.35$, for all linking lengths
 $L=$5, 7, and 10$\dist$.

 The triaxiality of superclusters can also be quantified using the
 measurement commonly employed for geometrical descriptions of
 elliptical galaxies, first introduced explicitly by \citet{statler94}:
 $$
 T \equiv \frac{1-(a_2/a_1)^2}{1-(a_3/a_1)^2}
 \eqno(3)
 $$
 With $a_1 \ge a_2 \ge a_3$, $T$ will take on values between 0 and 1
 (unless $a_1 = a_2 = a_3$, in which case $T$ is undefined).
 Two-dimensional pancake structures approach $T\to0$, while filaments
 approach $T\to1$ (note, however, that the increase is not linear with axis
 ratio).  Intermediate values of $T$ represent triaxial
 systems, with triaxiality increasing with $T$.  By this measure our
 low-$z$ superclusters are triaxial, ranging from a mean of
 $T \approx 0.65$ for $L = 10 \dist$ to $T \approx 0.69$ for
 $L = 5 \dist$.

 For all linking lengths we see a tail in the
 distribution that describes structures which are very nearly
 one-dimensional.  Thus, some mass concentrations are filamentary
 at $z\approx0$, especially in the relatively
 dense regions probed by $L = 5 \dist$.  Indeed, we would expect structures
 that have collapsed along two dimensions to be of higher average density
 than those that have collapsed along only one dimension.  This result
 agrees with HBB's identification of filamentary density enhancements in the
 large-scale structure; here we add the prediction of more diffuse
 flattened structures surrounding and possibly connecting the filaments.

\section{SUPERCLUSTER EVOLUTION}     \label{sec:evo}
 We examine the time evolution of the supercluster properties studied
 above.  We study superclusters in two redshift slices at
 $0.8 \le z \le 1.2$ and $1.7 \le z \le 2.4$, denoted
 $z \approx 1$ and $z \approx 2$, respectively.

 Superclusters are identified by the FOF method as before.  However, since
 the cluster abundance decreases at higher redshift, and their mean
 separation thus increases, we compare superclusters found at different
 redshifts {\em not} with the same linking length $L$, but instead with
 the same linkage parameter $b \equiv L/r_{av}$, where $r_{av}$ is the mean
 comoving intercluster separation.  As discussed in \S2.2, this approach
 sets a minimum overdensity for detection of superclusters, which is constant
 with redshift.

 This use of a constant linkage parameter accounts for the reduced density
 of clusters at high redshift.  However, because clusters are clustered,
 the density of superclusters depends on the relative rates of cluster
 formation in regions of high density vs. low density.  Hierarchical
 supercluster formation, in which clusters gravitate toward each other over
 time, also contributes to a reduced supercluster density at early times.

 We indeed find a lower density of superclusters in the high-$z$ cone.
 The total number of superclusters decreases by a factor of a few from
 $z \approx 0$ to $z \approx 1$, and by over an order of magnitude
 from $z \approx 1$ to $z \approx 2$.
 In order to construct a sufficiently large sample, we consider
 broad ranges in redshift, especially for the $z \approx 2$ slice.  In
 addition, we omit the $b$-value corresponding to the $L = 3 \dist$
 ($z \approx 0$)
 case from our subsequent analysis, as it results in too few superclusters
 at high redshift.

 As stated in \S2.3, our $z \approx 0$ linking lengths of
 $L = 5,7,10 \dist$ correspond to $b = 0.23,0.33,0.47$,
 respectively.  We use these latter linkage parameters $b$ to refer to the
 linking lengths used in our subsequent analysis.
 A given value of the $b$-parameter can be converted to an actual comoving
 linking length at a given redshift simply by multiplying by the mean
 intercluster separation at that redshift.
 The mean cluster densities in our $z \approx 1$ and $z \approx 2$ shells
 are $n_{av} = 3.53\times10^{-5}$ and $2.57\times10^{-6} (\dist)^{-3}$,
 corresponding to mean cluster separations of $r_{av} = 30.5$
 and $73.0 \dist$, respectively.  Therefore, $b=$ 0.23, 0.33, and 0.47
 correspond to $L=$ 7, 10, and 14.3$\dist$ at $z\approx1$, and 16.8, 24,
 and 34.3$\dist$ at $z\approx2$.

\subsection{Multiplicity Function}     \label{sec:mevo}
 The evolution of the multiplicity function is presented in 
 Fig$.$\,\ref{fig10}.
 We present the integrated multiplicity function for all three redshifts,
 using two different values of $b$.  The number density of superclusters drops
 noticeably with increasing redshift.  Monte Carlo simulations analogous
 to those described in \S2.3 were also run at $z \approx 1$ and $2$, with
 results similar to those obtained at low $z$: the multiplicity functions
 are physically significant for all multiplicities and linking lengths,
 except for binary clusters in the $b = 0.47$ case,
 where only $\sim 30\%$ more binary clusters are found over the Monte Carlo
 results.

 The redshift evolution of the multiplicity function depends on the
 linking length of the superclusters.  The vertical distances between
 the $b=0.47$ curves in Fig$.$\,\ref{fig10}
 increase with increasing multiplicity,
 implying that the abundance of the richest (highest-$N$) superclusters
 found with this linking length drops off especially swiftly with
 increasing $z$.  However, for $b = 0.23$ the vertical
 distances {\em decrease} with increasing $N$, to the extent that the
 abundance of the richest superclusters found with this linking length
 is in fact roughly the same at $z \approx 1$ as it is today.  We interpret
 these trends as further evidence that small linking lengths (e.g.
 $b = 0.23$) select the densest regions of
 superclusters.  Whereas the larger linking lengths do not find many
 rich superclusters at high redshift because of the
 lack of clusters, the smaller linking lengths find the same dense
 supercluster cores even at much earlier times.  These dense cores are
 likely the sites of the largest overdensities present at early times, and
 the first places where observable structures form through gravitational
 collapse.  Thus smaller linking lengths relative to the mean intercluster
 separation are needed to locate the highest overdensities in the
 observable Universe.  The lack of rich superclusters at high $z$
 is also evidence that these structures form largely through
 gravitational accretion of clusters that are initially too distant to be
 identified as supercluster members.

 Supercluster size remains highly correlated with multiplicity at high $z$,
 but with a lower $N-R_{max}$ normalization than at low $z$.  The size function
 does not evolve significantly with redshift for the larger superclusters
 (see Fig$.$\,\ref{fig11}), 
 while, as expected, the number of cluster members decreases.

\subsection{Dimensionality}     \label{sec:dimen}
 Using the method described in \S\ref{sec:shapes}, we fit ellipsoids to the
 superclusters identified at $z \approx 1$ and $z \approx 2$.  Because we
 only perform these fits on superclusters with $N \ge 5$, which are rare at
 high redshift, the statistical uncertainties are larger.  For example, only
 68 superclusters are identified at $z \approx 2$ for $b = 0.23$.

\begin{figure*}
\resizebox{\hsize}{!}{\includegraphics[angle=0]{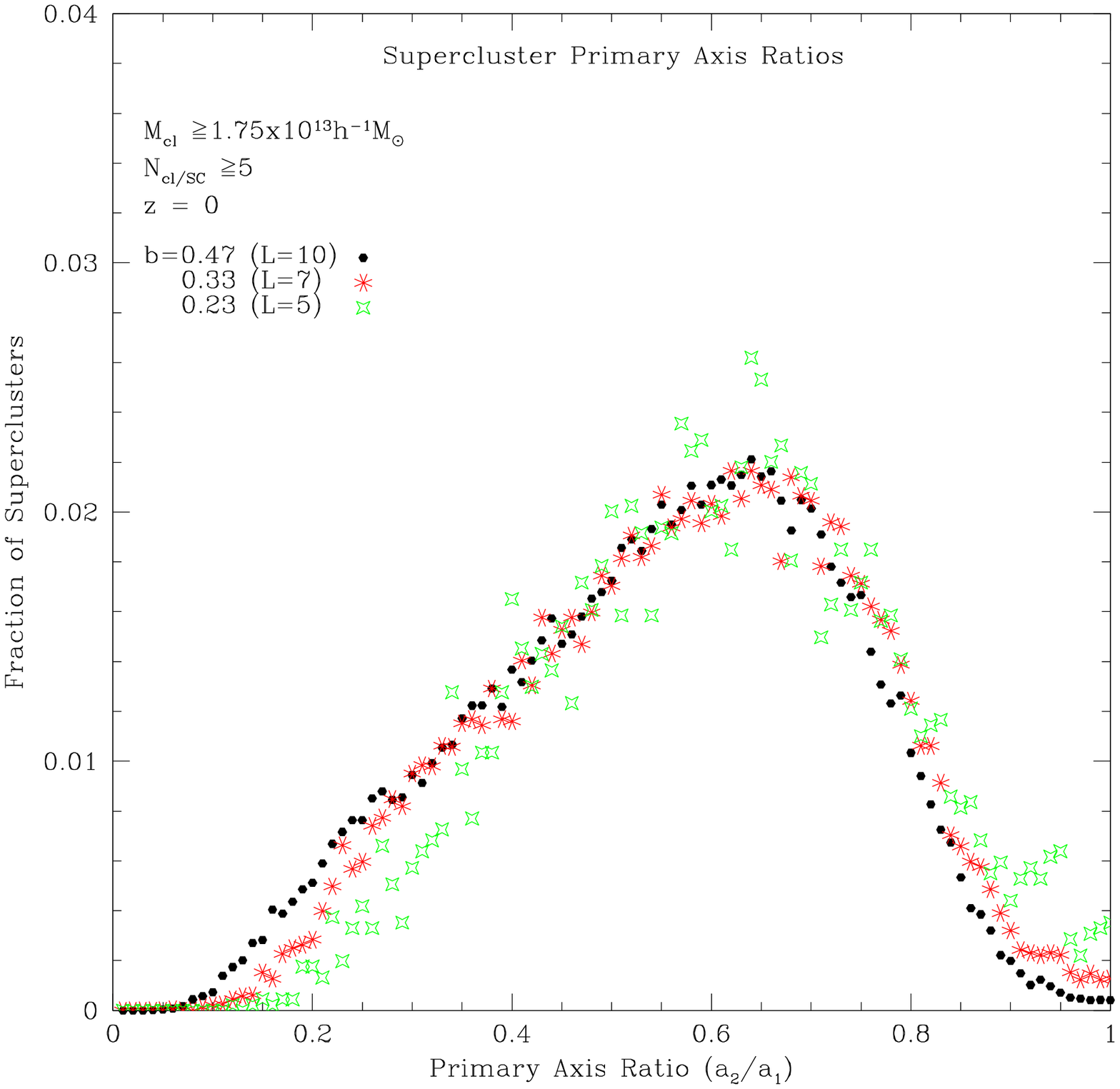}}
\caption{\label{fig6}
 Distributions of supercluster primary axis ratios for various linking
 lengths, at $z \approx 0$.  Linking lengths $L$ are quoted in $\dist$.
}
\end{figure*}

\begin{figure*}
\resizebox{\hsize}{!}{\includegraphics[angle=0]{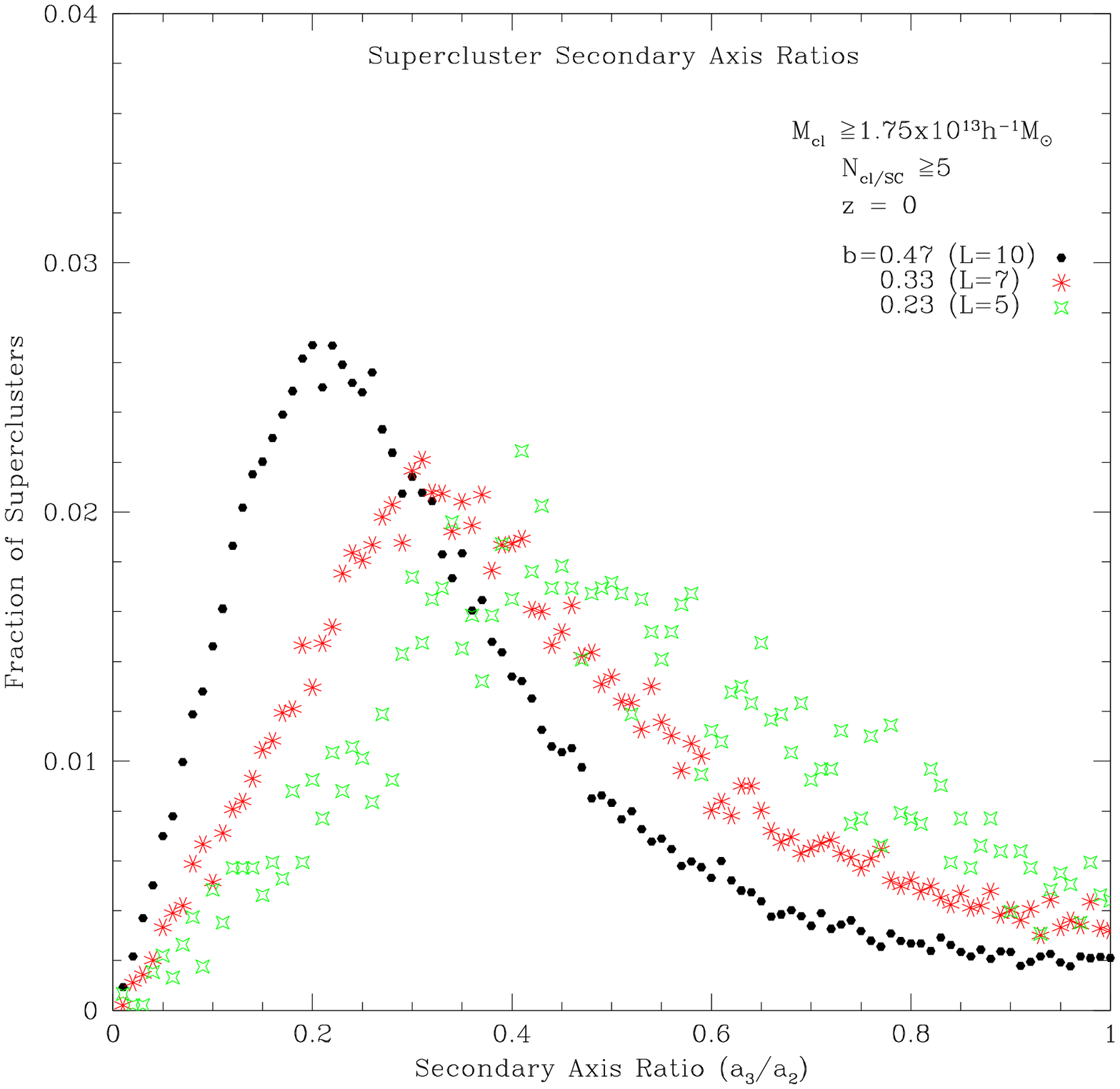}}
\caption{\label{fig7}
 Distributions of supercluster secondary axis ratios for various linking
 lengths, at $z \approx 0$.  Linking lengths $L$ are quoted in $\dist$.
}
\end{figure*}

\begin{figure*}
\resizebox{\hsize}{!}{\includegraphics[angle=0]{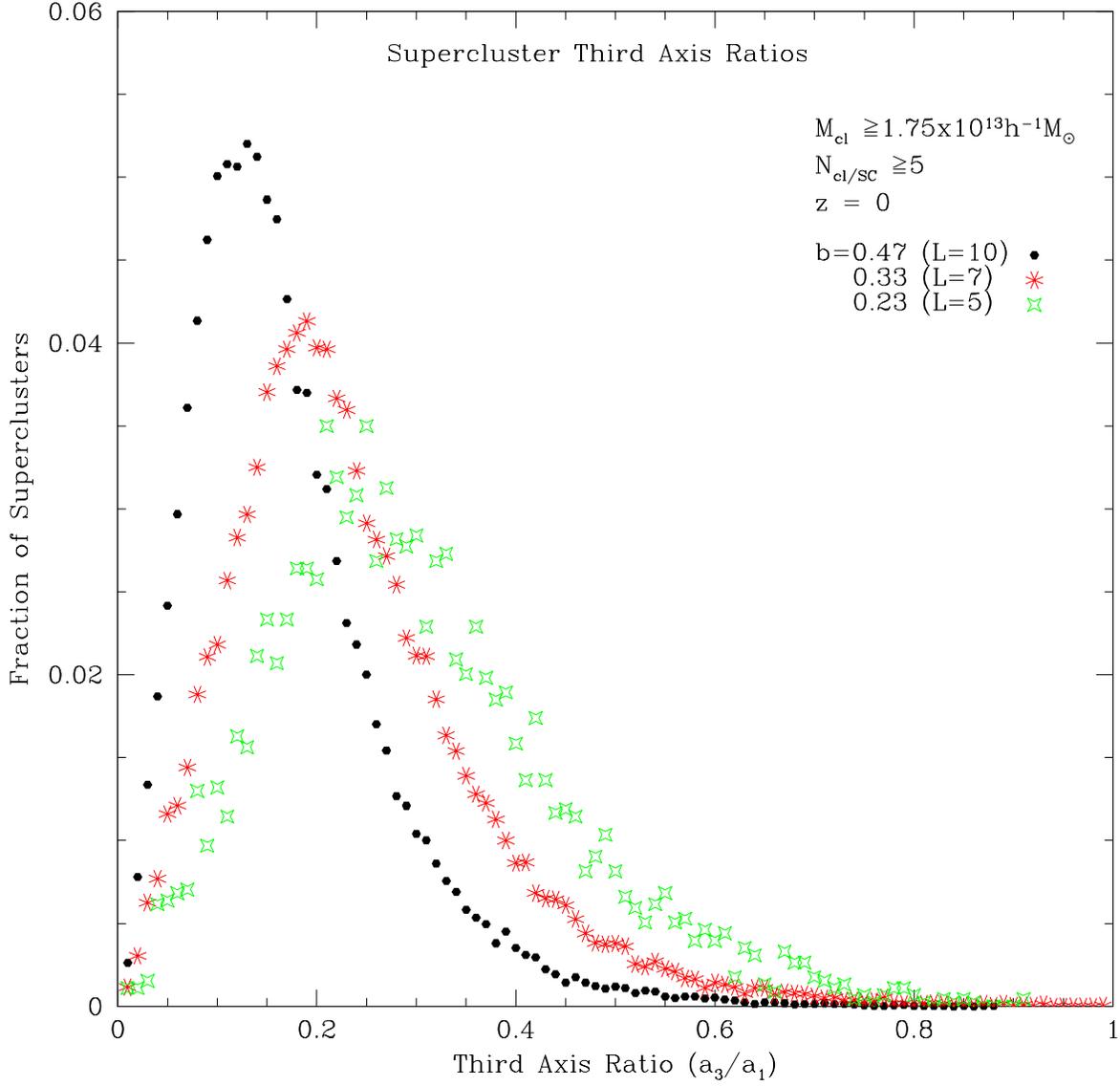}}
\caption{\label{fig8}
 Distributions of supercluster third axis ratios $a_3/a_1$
 for various linking lengths, at $z \approx 0$.
 Linking lengths $L$ are quoted in $\dist$.
}
\end{figure*}

\begin{figure*}
\resizebox{\hsize}{!}{\includegraphics[angle=0]{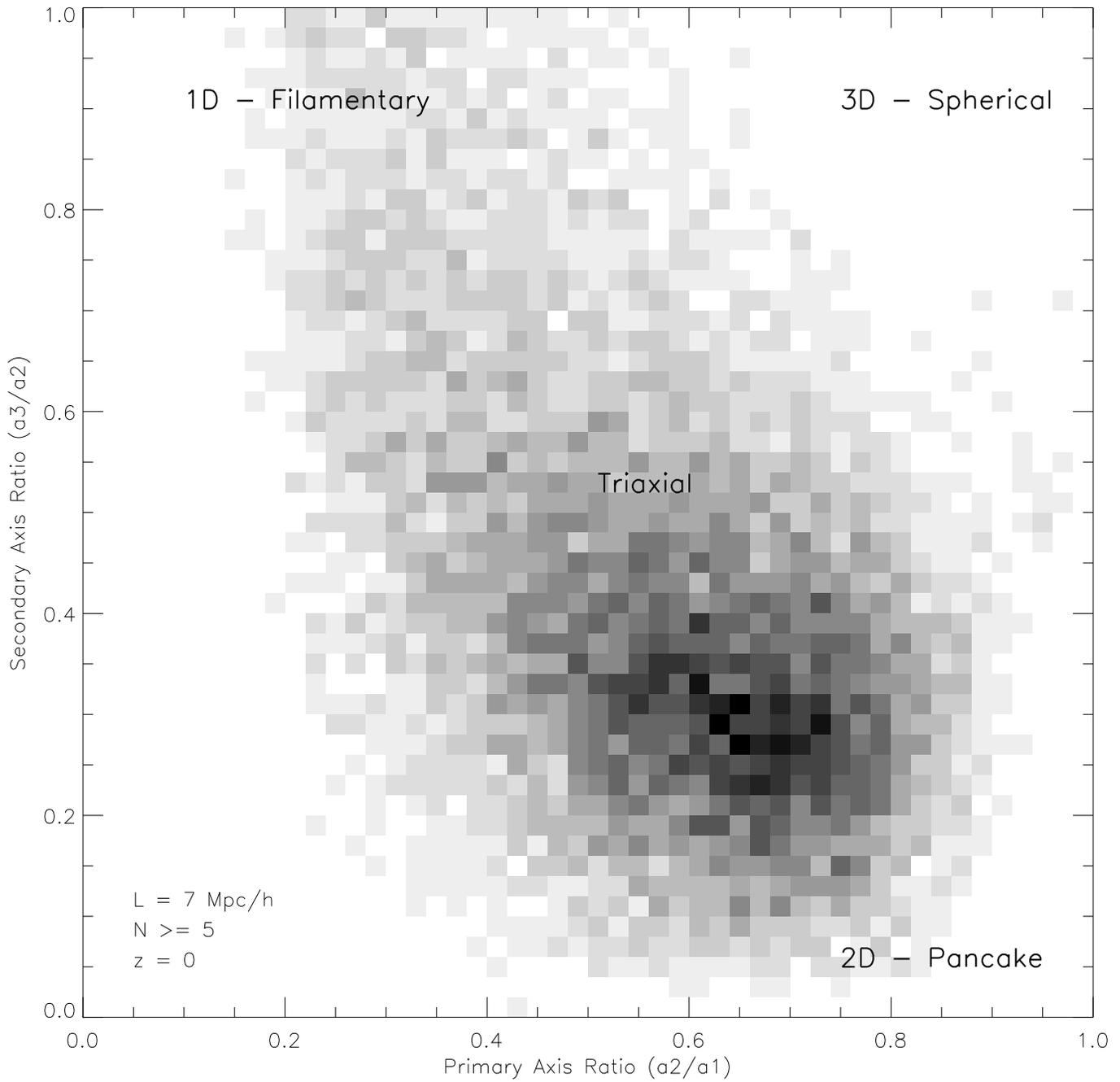}}
\vskip+0.5in
\caption{\label{fig9}
 Bivariate distribution of primary and secondary axis ratios for low-$z$
 superclusters found with linking length $L = 7 \dist$.  See the text
 (\S\ref{sec:shapes}) for interpretation.
}
\end{figure*}

\begin{figure*}
\resizebox{\hsize}{!}{\includegraphics[angle=0]{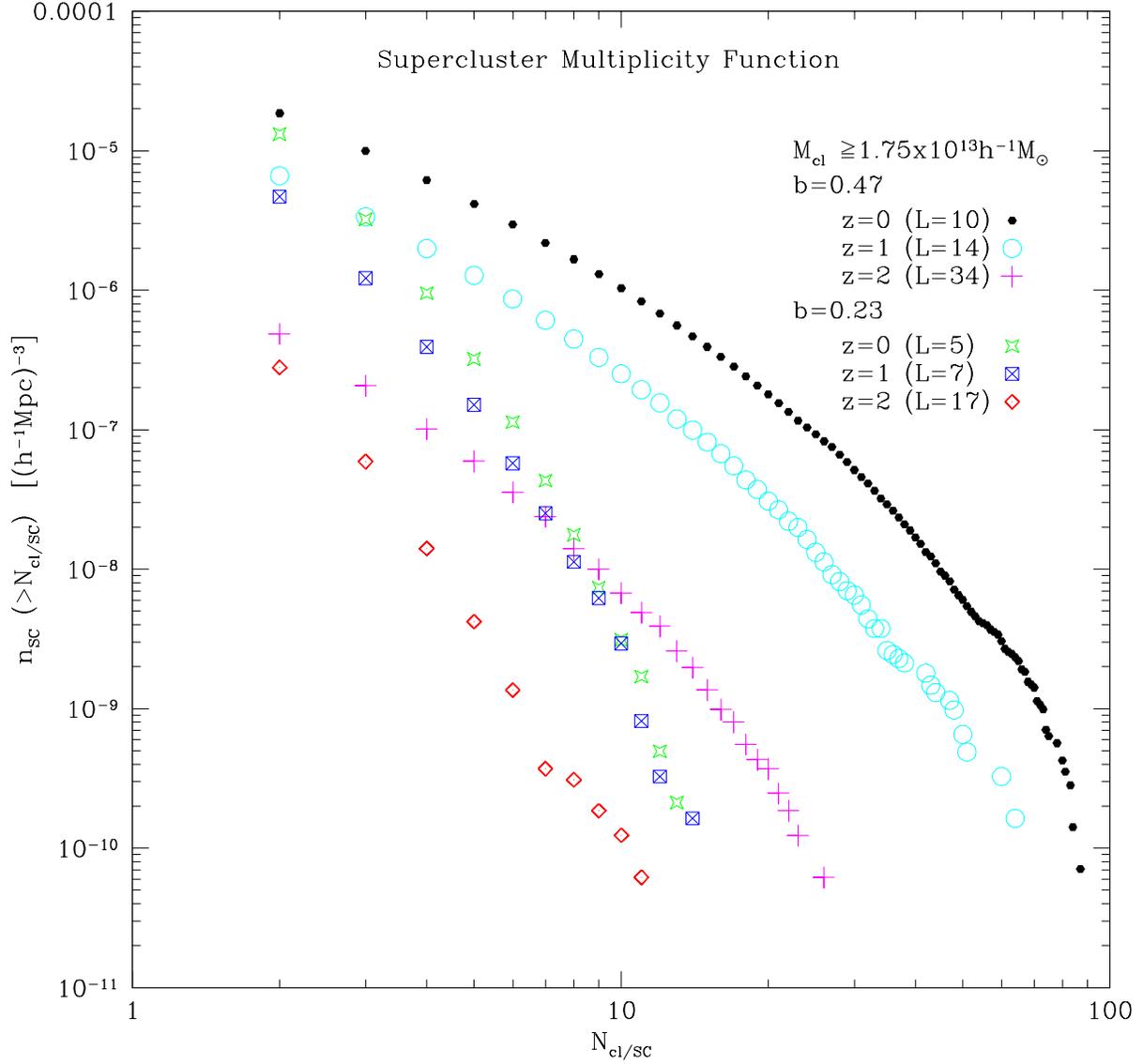}}
\caption{\label{fig10}
 Evolution of the integrated supercluster multiplicity function with redshift.
 Linking lengths $L$ are quoted in $\dist$.
}
\end{figure*}

\begin{figure*}
\resizebox{\hsize}{!}{\includegraphics[angle=0]{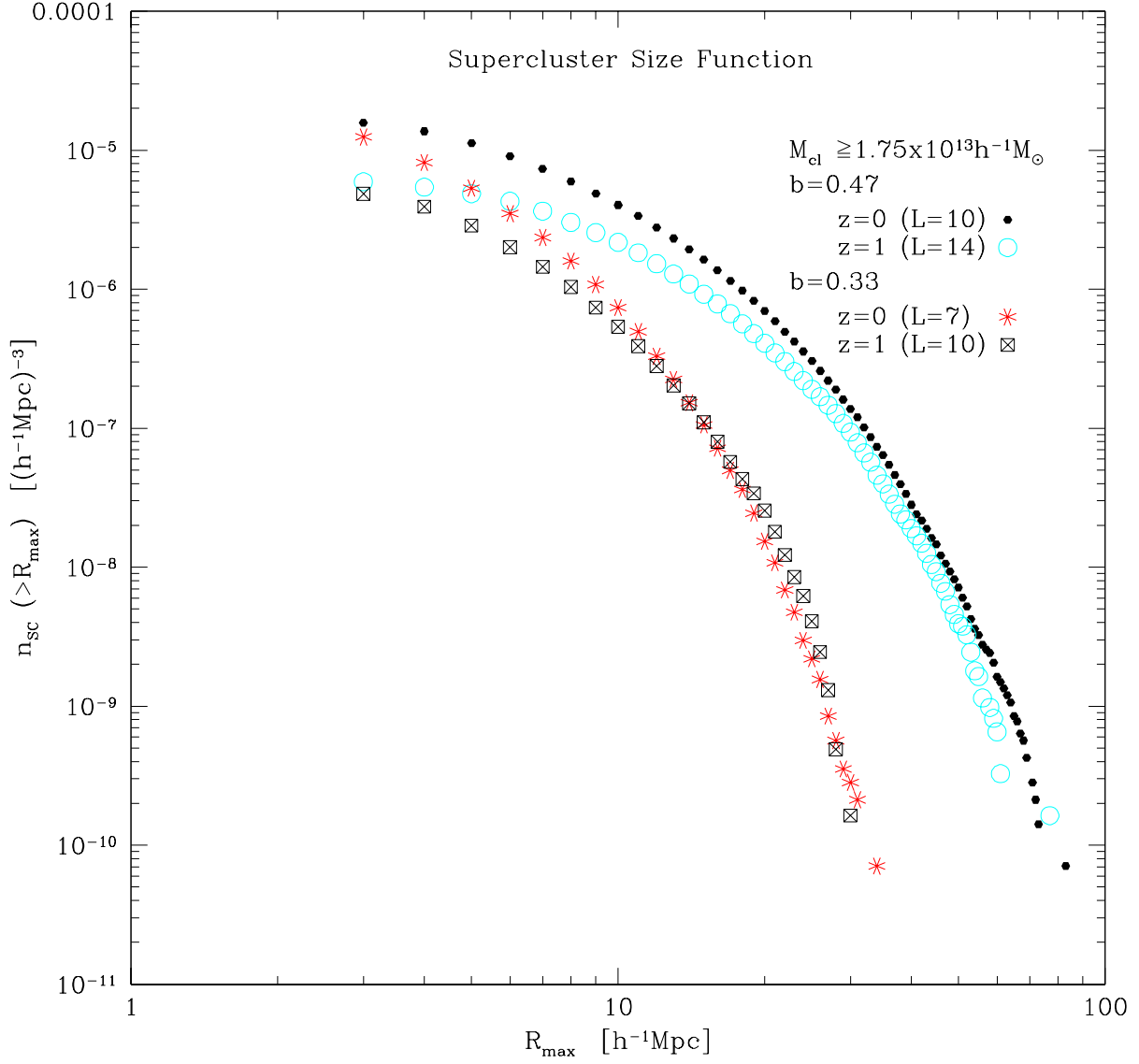}}
\caption{\label{fig11}
 Evolution of the integrated supercluster size function (comoving scales)
 with redshift.  Linking lengths $L$ are quoted in $\dist$.
}
\end{figure*}

\begin{figure*}
\resizebox{\hsize}{!}{\includegraphics[angle=0]{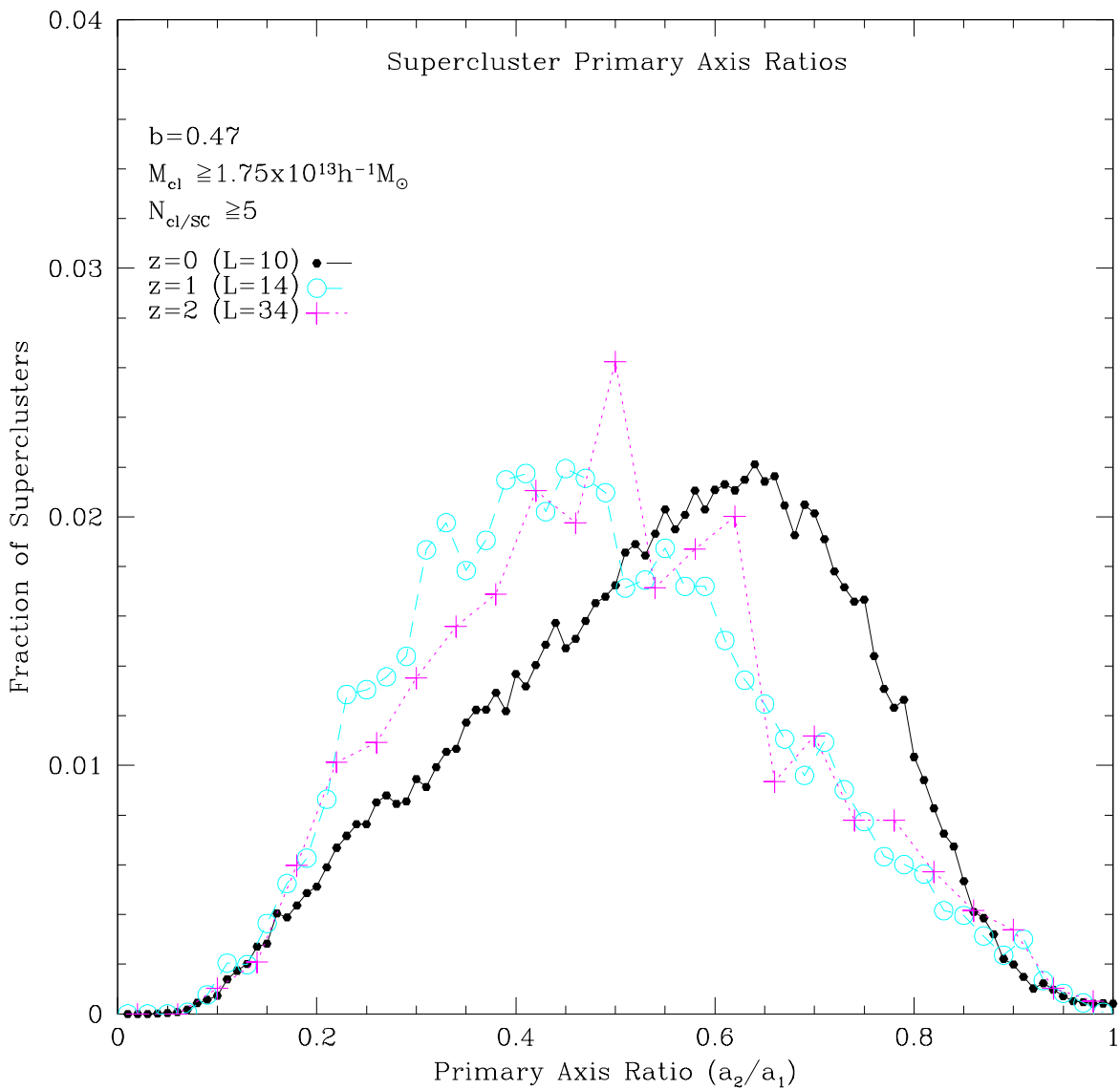}}
\caption{\label{fig12}
 Evolution of supercluster primary axis ratios with redshift.  All curves
 are generated from superclusters selected with linking length parameter
 $b = 0.47$.  The points are connected for clarity.
 Linking lengths $L$ are quoted in $\dist$.
}
\end{figure*}

\begin{figure*}
\resizebox{\hsize}{!}{\includegraphics[angle=0]{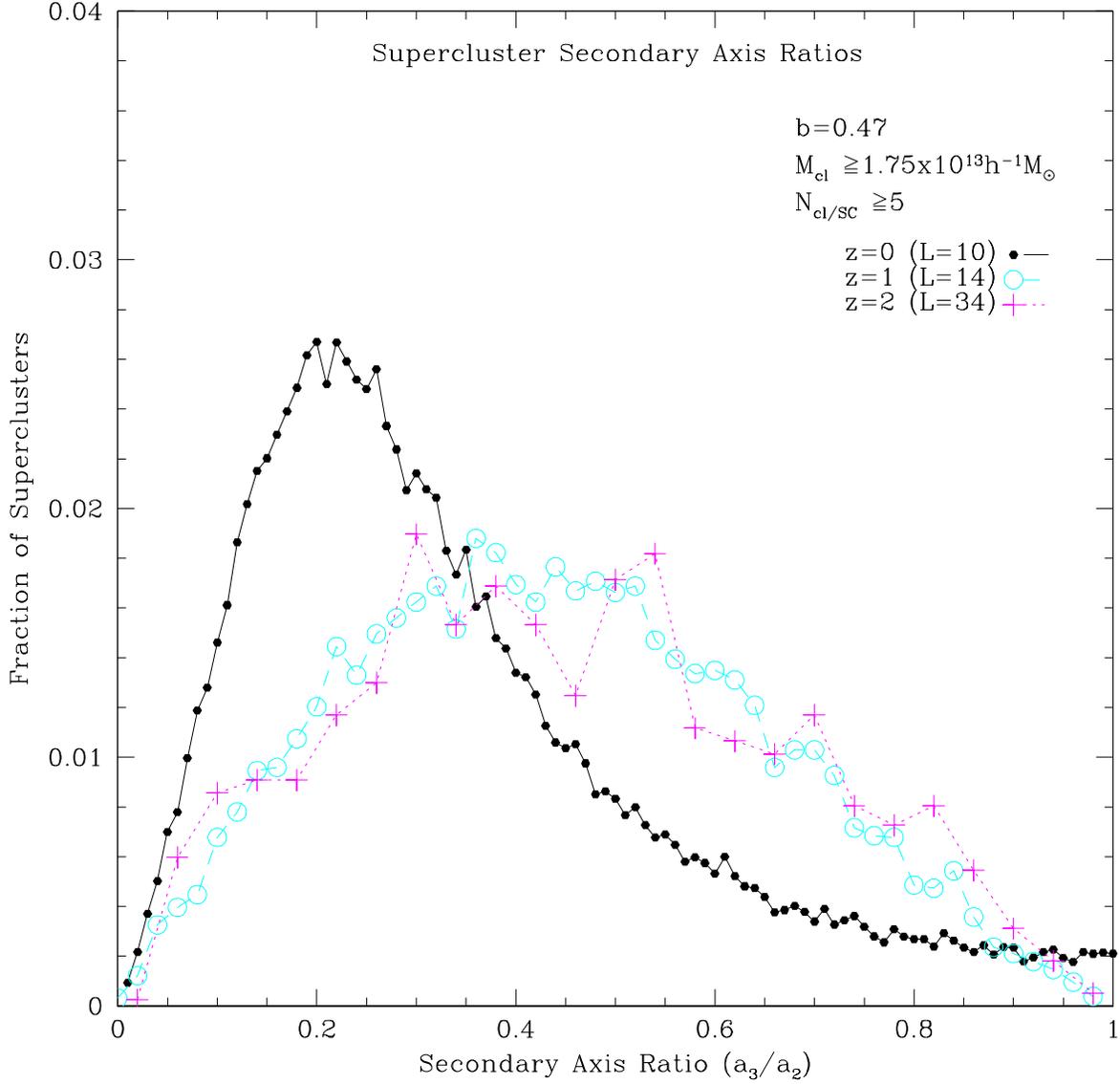}}
\caption{\label{fig13}
 Evolution of supercluster secondary axis ratios with redshift.  All
 curves are generated from superclusters selected with linking length
 parameter $b = 0.47$.  Linking lengths $L$ are quoted in $\dist$.
}
\end{figure*}

 We use the axis ratios of the best-fit ellipsoids to characterize the
 evolution of supercluster dimensionality.  Tables 2 and 3 present the
 mean primary and secondary axis ratios, respectively, for
 $b = 0.23,0.33,0.47$ at $z \approx 0,1,2$.  We also construct the
 axis ratio distributions 
 (similar to Figs$.$\,\ref{fig6}-\ref{fig7}) as a function of $z$
 for $b = 0.33$ and $0.47$
 (there are too few $b = 0.23$
 superclusters at high redshift to create meaningful distributions).
 The results for $b=0.47$ are shown in
 Figs$.$\,\ref{fig12}-\ref{fig13}. 
 The number of high-redshift superclusters is too low to construct
 bivariate distributions as in Fig$.$\,\ref{fig9},
 but we can consider how the changes in {\em mean} ratios correspond to
 translations of these distributions.

 The strongest evolution observed is the increase in the secondary
 axis ratio with increasing redshift, for $b = 0.47$ (Fig$.$\,\ref{fig13}).
 Most of the change occurs between $z \approx 0$ and $z \approx 1$, with
 only a slight additional increase between $z \approx 1$ and $z \approx 2$
 (see Table 3).  Typical secondary axis ratios increase from
 $a_3/a_2 \approx 0.25$ at low $z$ to $a_3/a_2 \approx 0.45$ at high $z$.
 Thus the flattened structures observed with $b = 0.47$ at $z \approx 0$
 (see \S\ref{sec:shapes}) 
 appear less flattened at high redshifts, indicating their
 gravitational collapse over time. 
 Very little evolution is evident
 in the secondary axis ratio for superclusters identified with $b = 0.33$,
 but the population of high-density filaments with $a_3/a_2 \approx 1$
 evolves between $z \approx 1$ and $z \approx 0$, as there are almost none
 of these filamentary structures at high redshift.
 The $z=0$ distributions for $b=0.33$ are shown in
 Figs$.$\,\ref{fig6}-\ref{fig7}; at higher redshifts the distributions
 are similar to the corresponding curves in 
 Figs$.$\,\ref{fig12}-\ref{fig13}. 

 The primary axis ratios are slightly smaller at higher redshifts
 for both $b = 0.33$ and $0.47$.  Typical values decrease from
 $a_2/a_1 \approx 0.6$ at $z \approx 0$ to $a_2/a_1 \approx 0.5$ at
 $z \approx$ 1-2.  In a plot similar to Fig$.$\,\ref{fig9}, 
 this would correspond
 to a leftward shift in the distribution, i.e. a shift toward greater
 triaxiality at high redshift, and a shift to a 2-D geometry as the systems
 approach $z \approx 0$.  Quantitatively, our high-$z$ superclusters have
 mean triaxiality $T \approx 0.79$ for both linking lengths, significantly
 higher than the mean $T$ for any linking length at low $z$.

 We conclude that superclusters had highly triaxial structure at early
 times, and that between a cosmological look-back time of $\sim 8$ Gyr
 (corresponding to $z \approx 1$ for the cosmology described in \S2.1) and
 the present day, many collapsed along their smallest dimensions to form the
 more nearly pancake-like structures observed in the simulation at
 $z \approx 0$.  In the same span of time, some structures in regions
 of highest density also collapsed along a second dimension to form
 nearly one-dimensional filaments.

\section{SUPERCLUSTERS SELECTED USING HIGH-MASS CLUSTERS} \label{sec:him}
 How do the supercluster properties change if superclusters are selected
 using higher-mass clusters as their seeds?  While the abundance of high-mass
 clusters is significantly lower, and the statistical uncertainties larger,
 the massive clusters are more easily detected in observational surveys.
 We repeat the analysis described above for superclusters selected using
 the rarer, higher-mass clusters with $M_{vir} \ge 1\times10^{14} \mass$.
 This threshold corresponds to typical ``rich clusters''
 \citep[e.g.][]{abell58, bahcall88}.  There are $9.68\times10^4$ clusters above
 this mass threshold at $z\approx0$, and $4.88\times10^3$ clusters at
 $z\approx1$.  The cluster abundance is $6.8\times10^{-6} (\dist)^{-3}$ at
 $z\approx0$, decreasing to $0.8\times10^{-6} (\dist)^{-3}$ at $z\approx1$.
 The intercluster mean separation is $52.7\dist$ at $z\approx0$, increasing
 to $107.9\dist$ at $z\approx1$.

 Superclusters are selected from these clusters at $z\approx0$ and
 $z\approx1$ using the linkage parameters $b=0.33$ and $0.47$; these
 correspond to $L=17.4$ and $24.8\dist$ respectively at $z\approx0$, and
 $L=35.6$ and $50.7\dist$ at $z\approx1$.  The large linking lengths $L$
 reflect the large mean intercluster separation of these massive clusters.
 We also include, for comparison, the smaller linking length of $L=10\dist$
 at $z\approx0$.
 There are fewer superclusters identified,
 and a smaller number of cluster members in each,
 as compared to those selected using the low-mass clusters.
 The overall structure of the superclusters is not significantly changed, as
 described below.

\begin{figure*}
\resizebox{\hsize}{!}{\includegraphics[angle=0]{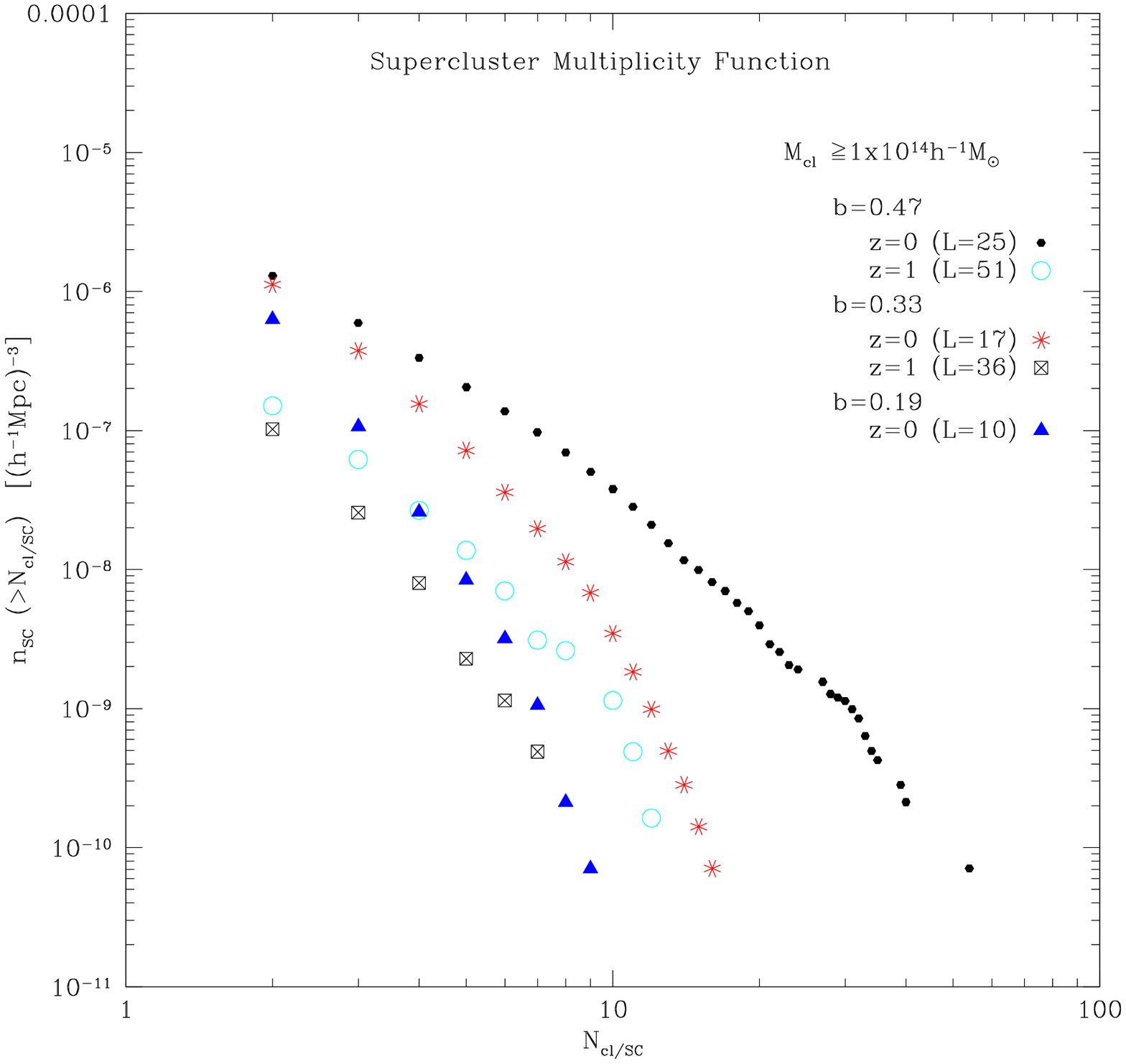}}
\caption{\label{fig14}
 Integrated supercluster multiplicity functions for superclusters selected
 using only high-mass clusters, $M_{vir}\ge10^{14}\mass$.
 Linking lengths $L$ are quoted in $\dist$.
}
\end{figure*}

 The multiplicity function of the superclusters is presented in 
 Fig$.$\,\ref{fig14}
 for both $z\approx0$ and $1$.  The results show, as expected, a similar
 shape to the multiplicity function of superclusters selected using the
 more numerous low-mass clusters 
 (Figs$.$\,\ref{fig3}, \ref{fig10}, \ref{fig11}), but with a greatly
 reduced amplitude: the number of superclusters at a given multiplicity is
 lower, and---equivalently---the supercluster multiplicity (richness) is
 smaller for a given supercluster abundance.  This is
 partly due to the much smaller number of massive clusters.  The
 evolution of the multiplicity function is stronger than that of
 superclusters selected with low-mass clusters (cf. Fig$.$\,\ref{fig10}).
 Superclusters with linkage parameters of $b\approx0.2 - 0.33$ contain up
 to $\sim10 - 15$ rich clusters at $z\approx0$.  The supercluster
 multiplicity decreases sharply at higher redshifts.

\begin{figure*}
\resizebox{\hsize}{!}{\includegraphics[angle=0]{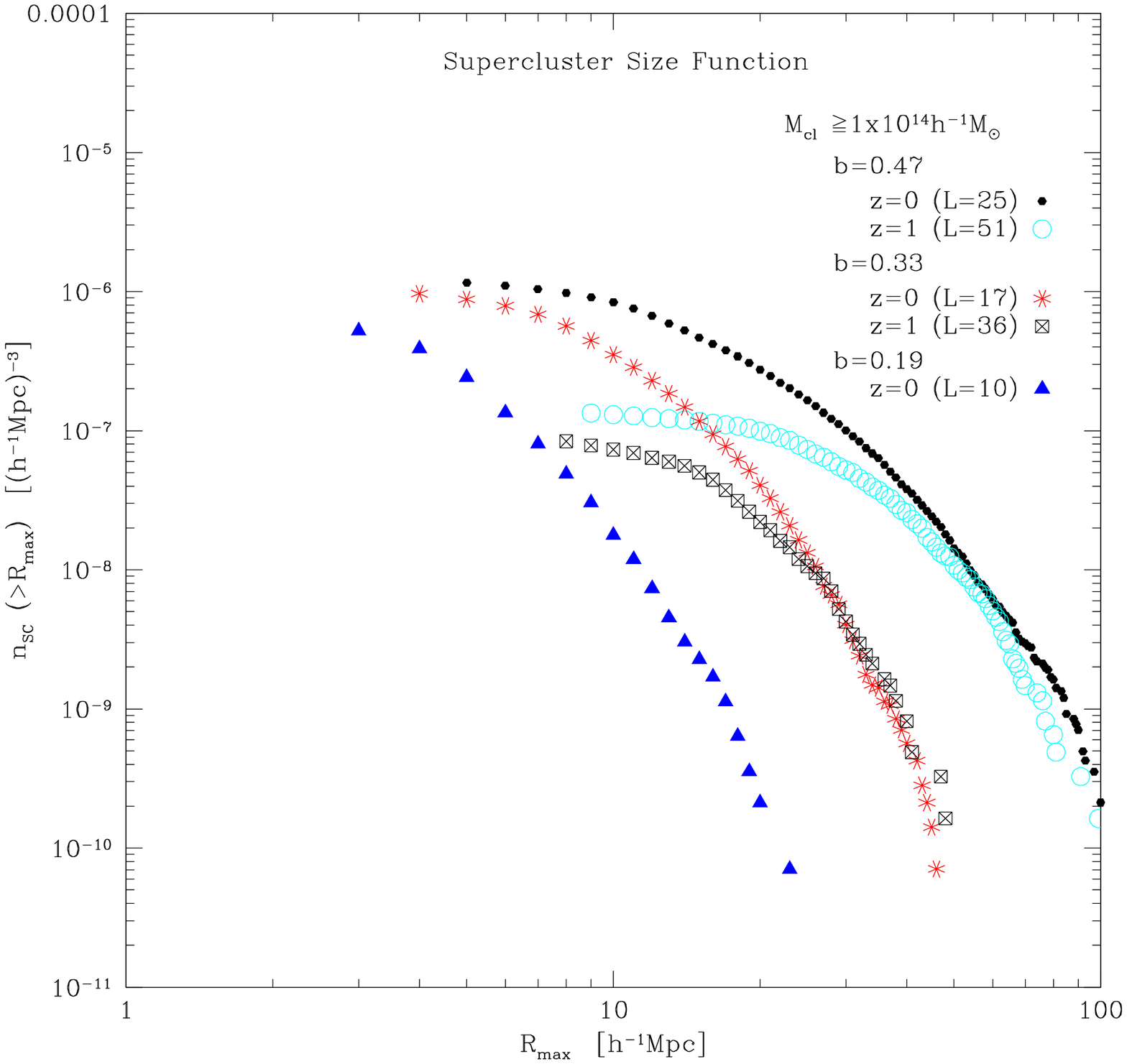}}
\caption{\label{fig15}
 Integrated supercluster size functions for superclusters selected
 using only high-mass clusters, $M_{vir}\ge10^{14}\mass$.
 Linking lengths $L$ are quoted in $\dist$.
}
\end{figure*}

 The supercluster size function is presented in Fig$.$\,\ref{fig15}.
 The maximum
 size of $z\approx0$ superclusters ranges from $\sim40\dist$ (i.e., twice
 the plotted $R_{max}$ radius) for $L=10\dist$, to $\sim90\dist$ for
 $L=17.4\dist$ ($b=0.33$), to $\sim200\dist$ for $L=24.8\dist$ ($b=0.47$).
 For a given linkage parameter $b$, the maximum supercluster size is
 somewhat larger than that of superclusters selected with low-mass clusters,
 since the linking length values $L$ are larger.  For a given value of $b$,
 no significant evolution is observed in the size of the
 largest superclusters, reflecting the existence of these extended
 structures at early times.  The same result is observed for superclusters
 selected with the lower mass threshold clusters.  The number of cluster
 members within these large structures decreases with redshift, as seen above,
 due to the lowered cluster abundance at high $z$.
 The supercluster multiplicity versus size relation 
 is similar---but reduced in amplitude, as expected---to the
 superclusters selected with the lower mass threshold 
 (Fig$.$\,\ref{fig5}).
 A nearly
 linear relation between multiplicity and size is observed for all but the
 binary clusters.

 The ellipticity distribution shows that superclusters
 exhibit a triaxial geometry on average, but with significantly larger
 scatter than the more numerous superclusters selected with lower-mass
 clusters (Figs$.$\,\ref{fig12}-\ref{fig13}).  
 The mean primary and secondary axis ratios are each
 $\sim0.4$ to $0.5$; for example, for $b=0.33$ at $z\approx0$, the mean
 values are $a_2/a_1 = 0.46 \pm 0.01$ and $a_3/a_2 = 0.40 \pm 0.01$.
 These ratios do not change significantly with either
 linking length or with redshift to $z\approx1$.  The distribution of
 ellipticities around this mean is, however, quite broad.  The supercluster
 shapes are not significantly different from those selected using the
 lower-mass clusters, except for the stronger flattening observed in the
 latter at $z\approx0$, where the mean triaxial shape has evolved slightly
 towards a flattened two-dimensional shape (with $a_2/a_1 \sim0.6$ and
 $a_3/a_2 \sim0.3 - 0.4$).  The superclusters selected with high-mass
 clusters are triaxial, with somewhat higher triaxiality at $z\approx0$ than
 those selected using lower-mass clusters---i.e., the
 high-mass superclusters are approaching a slightly more filamentary nature,
 with mean axis ratios of $\sim0.4$ in both $a_2/a_1$ and $a_3/a_2$ (cf.
 Fig$.$\,\ref{fig9}), and a mean triaxiality of $T=0.8 - 0.85$.

\section{CONCLUSIONS}    \label{sec:conc}
 We have investigated the multiplicities, sizes, radial
 structures, and dimensionalities of superclusters from $z \approx 0$ to
 $z \approx 2$ in $\Lambda$CDM cosmology.  The results 
 improve our understanding of the properties (especially shapes) and
 evolution of superclusters, and
 can also be used as
 predictions for testing the cosmology with future observations.

 At low redshift, we predict a total number density of
 $\sim4\times10^{-6}, 2\times10^{-6}$, and $3\times10^{-7} (\dist)^{-3}$
 superclusters with at least five member clusters with mass
 $M_{vir} \ge 1.75\times10^{13} \mass$ for superclusters selected
 with linking length $L = 10,7$, and $5 \dist$, respectively.  The maximum
 size of these superclusters ranges from $\sim150\dist$ for $L=10\dist$
 to $\sim30\dist$ for $L=5\dist$.  The
 abundance of superclusters decreases rapidly with increasing redshift.
 There are many more superclusters at low $z$ than at high $z$ largely
 because there are many more clusters that have formed by the present
 time and that have gravitated to form superclusters.
 The richest superclusters are almost entirely missing at high
 redshifts; these structures form late in cosmic time.  However, the densest
 ``core'' regions of superclusters, traced well by FOF linking
 lengths $\sim25\%$ of the mean intercluster separation, are present
 at $z\approx1$ almost as abundantly as they are today.  These are likely
 the sites of the highest overdensities in the early Universe.

 The spherically averaged density profiles of
 superclusters are well fit by an isothermal profile, $n(r) \sim r^{-2}$,
 over a broad range in radius.  The profile is shallower at small radii
 and is steeper at large radii.  A nearly linear relation exists between
 supercluster size and multiplicity, arising from this density profile.

 The clusters residing in superclusters are
 more massive on average than un-clustered clusters, and clusters residing
 in rich superclusters are more massive than those in poorer
 superclusters.

 We find that superclusters are triaxial in shape, especially at early
 times, where the mean triaxiality of our sample is $T \approx 0.79$.
 Over time, many collapse along a single dimension to approach
 two-dimensional shapes on the largest scales (i.e., those sampled by
 the FOF linking length $L = 10 \dist$ at $z \approx 0$, using low-mass
 clusters with $M_{vir} \ge 1.75\times10^{13} \mass$); their cores
 remain triaxial.  Quantitatively, the mean
 triaxiality for long linking lengths decreases to $T \approx 0.65$ at
 $z \approx 0$, whereas for smaller $L$ (i.e., in the cores) it decreases
 to $T \approx 0.69$.  The mean primary axis ratio of superclusters
 at $z \approx 0$ is $\sim0.6$, corresponding to a primary ellipticity
 ${\epsilon}_1 \approx 0.4$.  Typical secondary axis ratios at $z \approx 0$
 are $a_2/a_1 \approx 0.45$ and $a_3/a_1 \approx 0.25$ for $L = 5 \dist$,
 and $a_2/a_1 \approx 0.25$ and $a_3/a_1 \approx 0.12$ for
 $L = 10 \dist$.  Gravitational collapse increases ellipticity, as the
 potential gradient is larger along the minor rather than the major axis
 \citep{lin65}; collapse thus leads to lower dimensionality
 \citep[cf.][for a more recent treatment]{yoshisato06}.  A nonnegligible
 population of dense, filamentary superstructures is also present at
 $z \approx 0$, though not at higher redshift.

 Superclusters selected using higher-mass clusters,
 $M_{vir} \ge 10^{14} \mass$, show consistent results, with a reduced
 multiplicity function, as expected.  These superclusters are triaxial in
 shape, with somewhat greater mean triaxiality at $z\approx0$
 ($T\approx$ 0.8-0.85) than superclusters with lower-mass clusters.

 These properties, derived from a large-scale, high-resolution cosmological
 simulation, provide 
 direct information on the size, content, shape, and evolution
 of superclusters, as well as
 testable predictions of the currently favored
 $\Lambda$CDM model with which future and currently accumulating observations
 can be compared.  Once SDSS is complete, the best statistical analyses to
 date of large-scale
 structure using observations will be possible, especially if accurate and
 well-behaved photo-$z$ algorithms are developed.  There are other
 exciting observing programs on the horizon.  The Pan-STARRS Project
 \citep{kaiser02} is expected to see first light in 2006, and will carry out
 multi-band imaging surveys to identify galaxy clusters, as well as
 gravitational weak lensing observations to map the matter distribution in
 the Universe.  The Large Synoptic Survey Telescope \citep{tyson02}, which
 is expected to see first light in 2012, will provide improved
 weak lensing maps that will probe the dark matter distribution to the
 largest scales.  The analysis presented here will aid in the
 comparison of these major observational data with the predictions of
 the concordance cosmological model.

 This research was supported in part by NSF grant AST-0407305.
 Computations were performed on the National Science Foundation
 Terascale Computing System at the Pittsburgh Supercomputing Center;
 additional computational facilities at Princeton were provided by
 NSF grant AST-0216105.

\clearpage
\begin{table*}
\caption{\label{tab1}
 Supercluster Axis Ratios at $z \approx 0$, Cluster Masses $M_{vir} \ge 1.75\times10^{13} \mass$
}
\begin{center}
\begin{tabular}{rrrrrrr} \tableline \tableline
$L$ & Mean $a_2/a_1$  & Peak $a_2/a_1$ & Mean $a_3/a_2$  & Peak $a_3/a_2$ & Mean $a_3/a_1$  & Peak $a_3/a_1$ \\ \tableline
5   & 0.60 $\pm$ .009 &  0.64          & 0.50 $\pm$ .007 &  0.41          & 0.29 $\pm$ .004 &  0.23          \\       
7   & 0.57 $\pm$ .004 &  0.62          & 0.43 $\pm$ .003 &  0.31          & 0.23 $\pm$ .001 &  0.19          \\
10  & 0.54 $\pm$ .002 &  0.64          & 0.33 $\pm$ .001 &  0.20          & 0.16 $\pm$ .001 &  0.13          \\ \tableline
\end{tabular}
\tablecomments{
 Peaks and mean values of the distributions shown in 
 Figs$.$\,\ref{fig6}-\ref{fig8}.  Statistical
 errors are given for the mean values.
}
\end{center}
\end{table*}

\begin{table*}
\caption{\label{tab2}
 Evolution of the Primary Axis Ratio $a_2/a_1$, $M_{vir} \ge 1.75\times10^{13} \mass$
}
\begin{center}
\begin{tabular}{rrrrrrr} \tableline \tableline
$z$ & $b=0.23$        & $b=0.33$        & $b=0.47$        \\ \tableline
0   & 0.60 $\pm$ .009 & 0.57 $\pm$ .004 & 0.54 $\pm$ .002 \\       
1   & 0.44 $\pm$ .03  & 0.47 $\pm$ .01  & 0.48 $\pm$ .01  \\       
2   & 0.52 $\pm$ .06  & 0.50 $\pm$ .03  & 0.50 $\pm$ .02  \\ \tableline
\end{tabular}
\tablecomments{
 Mean values are given as a function of the linking length parameter $b$.
 Statistical errors are given.  Note that $b=0.23$ corresponds to
 $L_{z=0}=5\dist$, $b=0.33$ to $L_0=7\dist$, and $b=0.47$ to $L_0=10\dist$.
}
\end{center}
\end{table*}

\begin{table*}
\caption{\label{tab3}
 Evolution of the Secondary Axis Ratio $a_3/a_2$, $M_{vir} \ge 1.75\times10^{13} \mass$
}
\begin{center}
\begin{tabular}{rrrrrrr} \tableline \tableline
$z$ & $b=0.23$        & $b=0.33$        & $b=0.47$        \\ \tableline
0   & 0.50 $\pm$ .007 & 0.43 $\pm$ .003 & 0.33 $\pm$ .001 \\       
1   & 0.36 $\pm$ .02  & 0.40 $\pm$ .01  & 0.44 $\pm$ .01  \\       
2   & 0.41 $\pm$ .05  & 0.41 $\pm$ .03  & 0.46 $\pm$ .01  \\ \tableline
\end{tabular}
\tablecomments{
 Mean values are given as a function of the linking length parameter $b$.
 Statistical errors are given.
}
\end{center}
\end{table*}

\clearpage

\clearpage

\clearpage

\clearpage

\clearpage

\clearpage

\clearpage

\clearpage

\clearpage

\clearpage

\clearpage

\clearpage

\clearpage

\clearpage

\end{document}